\documentclass[12pt]{iopart}
\usepackage{graphicx,psfrag,bm,color,subfigure}

\newcommand{\bx}{{\bf x}}
\newcommand{\be}{\begin{equation}}
\newcommand{\ee}{\end{equation}}
\newcommand{\beq}{\begin{eqnarray}}
\newcommand{\eeq}{\end{eqnarray}}
\newcommand{\ba}{\begin{array}}
\newcommand{\ea}{\end{array}}
\newcommand{\bea}{\begin{eqnarray}}
\newcommand{\eea}{\end{eqnarray}}
\newcommand{\vp}{\varphi}
\newcommand{\eps}{\epsilon}

\newcommand{\g}{\gamma_{\perp}}
\newcommand{\gp}{\gamma_{\|}}
\newcommand{\om}{\omega}

\newcommand{\Ep}{E^+}
\newcommand{\Pp}{P^+}

\newcommand{\ex}[1]{\mbox{e}^{#1}}
\newcommand{\im}[1]{\mbox{Im}\left[#1\right]}

\newcommand{\psimu}{\Psi_\mu(\bx)}

\newcommand{\bxp}{{\bf x^\prime}}

\begin{document}
\title{Ab initio self-consistent laser theory and random lasers}

\author{Hakan E. T\"ureci$^1$, A. Douglas Stone$^{2}$, Li Ge$^2$, Stefan Rotter$^{3,2}$
  and Robert J. Tandy$^2$}
\address{$^1$ Institute for Quantum Electronics, ETH Z\"urich, 8093 Z\"urich, Switzerland}
\address{$^2$ Department of Applied Physics, Yale University, New Haven,
CT 06520, USA}
\address{$^3$ Institute for Theoretical Physics, Vienna University of
Technology, A-1040 Vienna, Austria, EU}
\ead{douglas.stone@yale.edu}
\begin{abstract}
We review our recent work leading to steady-state solutions of the semiclassical (Maxwell-Bloch) equations of a laser.  These are coupled non-linear partial differential equations in space and time which have previously been solved either by fully time-dependent numerical simulations or by using major approximations which neglect non-linear modal interactions and/or the openness of the laser system. We have found a time-independent technique for determining these stationary solutions which can treat lasers of arbitrary complexity and degree of openness. Our method has been shown to agree with time-dependent numerical solutions to high accuracy and has been applied to find the electric field patterns (lasing modes) of random lasers, which lack a laser cavity and are so strongly damped that the linear system has no detectable resonances.  Our work provides a link between an important non-linear wave system and the field of quantum/wave chaos in linear systems.
\end{abstract}

\maketitle

\section{Introduction: semiclassical laser theory and its limitations}

Lasers are among the most important non-linear systems in modern science and technology.  They take advantage of the quantum phenomenon of stimulated emission to create an amplifier for light, which becomes an oscillator emitting in one or several very small frequency windows above a series of thresholds.  Shortly after the demonstration of the first lasers in the early 1960's (following experiments and theory on the maser and proposals for the 
laser), standard theoretical descriptions were developed independently by Haken and Sauermann \cite{haken63} and Lamb and collaborators\cite{lamb64,LangSL73} which neglect the quantum fluctuations of the electromagnetic field, hence these are referred to as semiclassical theories.  Shortly thereafter a full quantum theory of the laser was developed by Haken \cite{haken_laserbook}, and Scully and coworkers \cite{ssl_laserphysics}, which builds on the semiclassical theories.  

The semiclassical theory describes the electric and magnetic fields using Maxwell's equations, which couple the fields to matter through the non-linear polarization of the gain medium, which is described by quantum equations of motion. This theory describes all of the characteristic non-linear phenomena observed in lasers, such as mode competition and selection, bistability, frequency and phase locking.  The semiclassical laser equations in their most common form describe a gain  medium of identical two-level ''atoms'' with energy level spacing $\hbar \omega_a$ and relaxation rate $\gp$ being pumped by an external energy source, $D_0$, contained in a cavity which can be described by a linear dielectric function, $\eps (\bx)$. This leads to a population inversion of the atoms, $D(\bx,t)$ which in the presence of an electric field creates a non-linear polarization of the atomic medium, $P(\bx,t)$, which itself is coupled non-linearly to the inversion through the electric field, $E(\bx,t)$.  The electric field and the non-linear polarization are related linearly through Maxwell's wave equation, although the polarization is implicitly a non-linear function of the electric field in a manner we will demonstrate explicitly below.  The induced polarization also relaxes at a rate $\g$ which is typically much greater than the rate $\gp$ at which the inversion relaxes. This is a key assumption in our approach, as well as in almost all of the work in the literature on multi-mode lasing.  

The resulting system of non-linear coupled partial differential equations for the three fields $E(\bx,t),P(\bx,t),D(\bx,t)$ are ($c=1$): 
\bea
\ddot{E}^+  &=& \frac{1}{\eps (\bx)}\nabla^2 {E}^+ - \frac{4\pi}{\eps (\bx)} \ddot{P}^+
\label{eqMB0}\\
\dot{P}^+ &=& -(i\om_a + \g) \Pp + \frac{g^2}{i\hbar} \Ep D  \label{eqMB1}\\
\dot{D} &=& \gp\left( D_0-D \right) - \frac{2}{i\hbar}\left( \Ep (\Pp)^* -
\Pp (\Ep)^* \right)\,. \label{eqMB2}
\eea
Here $g$ is the dipole matrix element of the atoms and the units for the pump are chosen so that $D_0$ is equal to the time-independent inversion (uniform in space) of the atomic system in the absence of an electric field.  The electric field and polarization are real functions (vector functions in general, but we assume a geometry where they can be treated as scalars).  In writing the equations above we have written these fields in the usual manner in terms of their  positive and negative frequency components, $E =E^+ + E^-$, $P = P^+ + P^-$, and then made the rotating wave approximation (RWA) in which the coupling terms correspond to driving far away from the atomic transition frequency are neglected.  This approximation is both very good under general conditions and not essential; it just simplifies the equations.  We have {\it not} made another widely used approximation, the slowly-varying envelope approximation.  In this approximation the fields are factorized as an oscillatory term at the atomic transition frequency multiplied by a slowly varying envelope function, whose second derivative is then neglected.  We have shown recently \cite{LiTST08} that this approximation is not very accurate for microcavities, and it provides no significant simplification in our approach.

Note that if in this non-linear system the electric field and polarization are zero at $t=0$ then they remain zero and the pump just finds its steady-state value $D (\bx,t) = D_0$.  This is because the semiclassical theory neglects spontaneous emission, in which the atom spontaneously relaxes from the upper level through the emission of light, generating a fluctuating electric field.  If spontaneous emission is not neglected, then the pump will amplify this emission leading to a broad enhancement of light emission near the atomic transition frequency.  At a series of thresholds for multi-mode lasing much narrower lasing lines emerge out of this broad peak of amplified stimulated emission.  This process is simplified in the semiclassical theory.

In the semiclassical theory if there is an initial pulse of electric field to start the non-linear interactions, then three situations can occur: 

\begin{enumerate}

\item
The pulse will decay as $t \to \infty$ and there is no field in the cavity in the presence of the pump.  In this case the pump rate is below the lasing threshold.  

\item
There is a transient regime of a fluctuating field after which the system self-organizes to oscillate and emit light at a discrete set of lasing frequencies.  When there is only one frequency we are in the single-mode regime; otherwise we have multi-mode lasing.  In the single-mode regime the inversion 
$D(\bx,t)$ will no longer be uniform in space, but will be independent of time.  In the multi-mode regime the inversion will in principle depend on both time and space.  The polarization will oscillate with the same frequencies as the field.
 
\item 
The fields never settle down to a multiperiodic time dependence and quasiperiodic or chaotic behaviour occurs.  We neglect this possibility in the approach given below.  Simulations of (\ref{eqMB0})--(\ref{eqMB2}) indicate this does not occur in the parameter range we are investigating \cite{LiTST08}.  It is well-known, however, that chaotic time-dependence can occur in multi-mode lasers in general \cite{haken_light2}.

\end{enumerate}

Above the first lasing threshold in the semiclassical theory the inversion becomes non-uniform in space and is reduced from its value $D_0$ in the absence of lasing emission. This changes the effective gain in the cavity in a space-dependent manner, making it more difficult for higher lasing modes to reach threshold. Also, the single mode that is lasing saturates as the pump is increased.  Thus the modes of a laser have an effectively infinite order non-linear interaction through the gain medium; this effect is called spatial hole-burning.  Unlike previous theories, our approach is able to treat this effect to all orders, in an approximation which is found to be excellent.  

The laser is an open system, and the electric field in a laser is being generated by the stimulated emission of light from the atomic medium in the absence of any input light at the lasing frequencies.  As a consequence the stationary electric field in the laser cavity does not conserve energy; the flow of light energy (magnitude of the Poynting vector) is increasing along the cavity in the direction of the output(s).  Therefore the electric field in the cavity has a non-hermitian character  and we can show that it is determined by a non-hermitian boundary condition; outside the cavity there is no gain medium and energy flux is conserved.  This non-hermitian character of the actual lasing modes has been widely neglected in semiclassical laser theory which has until very recently only dealt with hermitian closed cavities in the multi-mode regime \cite{haken_light2}.  Heuristic arguments were used to discuss the output power of the laser and its emission pattern.  Modern nanofabrication capabilities along with potential applications have led to a wide variety of new microlasers with complex cavity structures; examples are  photonic bandgap defect mode lasers \cite{PainterLSYODK99}, deformed dielectric cavity lasers \cite{SchwefelTureci04} and random nano-composite lasers \cite{Wiersma08}.  For such lasers  the emission pattern and output power is not easily guessed from internal mode properties of the closed resonator.  Thus we sought a more rigorous and predictive framework for analyzing arbitrarily complex and open multi-mode lasers.  Our approach \cite{LiTST08,Tureci06}, which we call Ab Initio  Self-Consistent (AISC) laser theory  is able to treat lasers with any degree of openness, and as mentioned earlier, with arbitrarily strong multi-mode non-linear interactions.  Below we will review this new approach and describe a dramatic application to random lasers.  

Random lasers \cite{Wiersma08,Cao05} are aggregates of nanoparticles which have gain, or are embedded in a gain medium.  These lasers have no mirrors or reflecting boundaries to confine light; light generated in the gain medium by stimulated emission simply multiply-scatters as it diffuses out to the boundary and escapes from the aggregate.  We refer to these systems as diffusive random lasers (DRLs).  An original motivation for studying random lasers was the interest to possibly observe Anderson-localized states of the linear wave equation; however such localized-modes random lasers have only been demonstrated in one dimension \cite{Genack05}.  DRLs correspond to the limit of extreme openness where the system in the absence of gain has no long-lived resonances at all and hence no detectable natural frequencies to determine the oscillation frequencies of  the laser. Nonetheless random lasers behave in most respects like conventional multi-mode lasers, with sharp thresholds and line-narrowing above threshold.  Until our work there had been no solution of the semiclassical lasing equations for DRLs above threshold, except for numerical simulations of the equations in time and space which provides no physical picture of the lasing modes in such systems.  Below we show that our AISC laser theory is able to find and interpret the lasing modes in such a system, which have been beyond conventional treatments of the semiclassical lasing equations.

\section{Self-consistent  steady-state lasing equations}

The starting point of our new formulation is to assume that there exists a steady state multiperiodic solution of equations (\ref{eqMB0})-(\ref{eqMB2}) above, for long times after the pump is turned on, i.e., we try a solution of the form:
\be
\Ep (\bx, t) = \sum_{\mu=1}^N \psimu e^{-i k_\mu t},  \;\;\;\;\Pp (\bx, t) = \sum_{\mu=1}^N P_\mu (\bx) e^{-i k_\mu t}\,.
\label{eqEPansatz0}
\ee
Having taken  $c=1$ we do not distinguish between frequency and wavevector.  The functions $\psimu$ are the unknown lasing modes and the real numbers $k_{\mu}$ are the unknown lasing frequencies.  Unlike standard approaches, here these functions and frequencies are completely general and bear no {\it a priori} relationship to the normal modes or linear quasi-modes of the passive cavity (cavity without gain).  

Note from this ansatz that the non-linear term in (\ref{eqMB2}) for the inversion $\sim \Ep (P^{+})^*$ is time independent in steady-state for a single-mode solution ($N=1)$, allowing an exactly time independent inversion (which does vary in space).  However when there are two or more modes, the non-linear term in (\ref{eqMB2}) will now oscillate at the difference frequencies of the modes, ruling out a strictly stationary inversion in the multi-mode case.  This beating represents the interference of the different modes since they oscillate at different frequencies.  The other time scale in the dynamics of the inversion is $\gp^{-1}$; if this time scale is long compared to the beat periods of the modes, i.e., if $\gp << \Delta k_{\mu \nu} = |k_\mu - k_\nu|$, $\forall{\mu,\nu}$, then these time-varying interference terms average to zero.  In addition it must be assumed that the polarization follows the inversion closely, so $\g >> \gp$. In these two limits the inversion may be approximated as stationary to high accuracy.  This is a standard approximation, often used in multi-mode laser theory and often realized in lasers of interest; we will refer to it as the stationary inversion approximation (SIA). Once this approximation is made, then (\ref{eqMB1}) can be solved to give the polarization in terms of the lasing modes 
$\psimu$ and the unknown (stationary) inversion, $D_s(\bx)$.  Substitution of this result back into the equation for the inversion gives $D_s(\bx)$  in terms of $\{ \psimu  \}$ so that the polarization in (\ref{eqMB0}), which is the source term for the wave equation, is completely given in terms of $\{ \psimu  \}$.  The time derivatives needed in (\ref{eqMB0}) are determined by the multiperiodic ansatz (\ref{eqEPansatz0}) and one obtains an inhomogeneous differential equation in space only for $\Ep$.  This nominally inhomogeneous linear differential equation can be formally inverted with a Green function (determined by the outgoing wave boundary conditions) to yield a self-consistent equation determining all non-zero lasing modes.  The details of this have been given in \cite{LiTST08,Tureci06}; the resulting equation is:
\begin{equation}
\Psi_\mu (\bx) =  \frac{i \g}{\g - i (k_\mu -
k_a)}\frac{k_\mu^2}{k_a^2} \int_{\cal D} d {\bxp} \frac{D_0(\bxp)
G(\bx,\bxp;k_\mu)  \Psi_\mu (\bxp)}{\eps(\bxp) (1 + \sum_\nu
\Gamma_\nu |\Psi_\nu (\bxp)|^2)}\, . \label{eqTSG}
\end{equation}
Here $k_a$ is the frequency of the atomic transitions, $G(\bx,\bxp;k_\mu)$ is the Green function of the open cavity,
$\Gamma_\nu = \Gamma (k_\nu)$ is the gain profile evaluated at
$k_\nu$, $D_0(\bx)= D_0 (1 + d_0(\bx))$ is the pump, which we have now
allowed to vary in space, $\epsilon (\bx) = n^2(\bx)$ is the dielectric function of the ''cavity'', which can be arbitrarily complex or leaky.  This linear dielectric function of the cavity can also be frequency-dependent and complex, but we do not consider these cases here. Electric field and pump strength are measured in the natural dimensionless units of the problem, $e_c = \hbar \sqrt{\g \gp}/2g, D_{0c}= \hbar \g/4 \pi k_a^2 g^2$.  The integral in (\ref{eqTSG})
extends over the gain region, denoted by ${\cal D}$.

These are the fundamental equations of the AISC laser theory, a set of self-consistent non-linear integral equations for the unknown lasing modes and frequencies.  We have developed an efficient method for solving (\ref{eqTSG}) iteratively, which will be described below.  As expected, these equations exhibit a series of thresholds $D_{th}^{(i)}$ in the pump strength, $D_0$. $\Psi_\mu (\bx)=0$ is always a solution, but above some value of $D_0$ a first non-trivial solution appears.  It then interacts with itself through the non-linear hole-burning denominator in (\ref{eqTSG}) and at the same time partially suppresses other solutions which would turn on in the absence of the first mode.  As the pump increases additional solutions do appear in general, with those solutions favored which overlap less in space with the existing lasing mode(s).  Physically this is because gain is less depleted where the electric field of the existing lasing modes is small.  Note the presence of the unknown lasing frequencies both explicitly and implicitly in (\ref{eqTSG}).  They are determined by the requirement that the non-linear integral operator on the right hand side of (\ref{eqTSG}) have real eigenvalues for each solution, and the lasing frequency must be allowed to flow with each iteration of (\ref{eqTSG}) to maintain this ''gauge'' condition.  More details of this procedure will be given below. Despite the fact that this is a non-linear system, there is a natural basis set in which to expand the functions $\psimu$, which is determined by the Green function in (\ref{eqTSG}).  This basis set has the property that for a high-Q cavity each lasing mode become equal to a single basis function, as will be discussed in the next section.  In general (\ref{eqTSG}) then can be formulated as a non-linear map for the vector of coefficients in this basis set and the solutions are the fixed points of this map.

\section{Open cavity boundary conditions and CF states}

The Green function which appears in (\ref{eqTSG}), $G$, is that  of the cavity wave equation 
\be
[ \eps(\bx)^{-1}  \nabla^2  +  k^2 ]\,
G(\bx,\bx'; k) = \delta^3(\bx-\bx'),
\ee
with the real parameter $k=k_\mu$ being the unknown lasing frequency of mode $\mu$ in
(\ref{eqTSG}).  The crucial point in the above equation is the nature of the boundary conditions on $G$; at infinity only outgoing waves of frequency $k_\mu$ are allowed: $\nabla_r G(\bx,\bx'; k) = \nabla_{r \prime} G(\bx,\bx'; k) = ikG(\bx,\bx'; k)$, where 
$ \nabla_r$ is the radial derivative.   These are non-hermitian boundary conditions implying that the spectral representation of $G(\bx,\bxp; k)$ is of the form:
\begin{equation}
G(\bx,\bxp ;k) =  \sum_m
\frac{\varphi_{m}(\bx,k)\bar{\varphi}^*_{m}(\bxp,k)}{k^2 - k_m^2(k)}.
\label{eqspecrep2}
\end{equation}
We refer to the functions $\varphi_{m}(\bx,k)$  in (\ref{eqspecrep2}) as the constant-flux (CF) states, as they conserve the photon flux outside the cavity.  They satisfy 
\be
-  \eps(\bx)^{-1}  \nabla^2 \varphi_m(\bx,k) =  k_m^2(k)
\varphi_m(\bx,k)
\label{eqcf}
\ee
with the corresponding non-hermitian boundary condition of purely outgoing spherical waves at infinity. The non-hermitian boundary conditions imply that the CF wavevectors $k_m$ are always complex, and in fact it can be  shown that they always have negative imaginary part, corresponding to amplification within the cavity \cite{Tureci06}.
For all the cases considered here it is sufficient to impose the outgoing wave boundary condition at the boundary of
the cavity.  In fact the Green function only correctly calculates the electric field inside the cavity.  By connecting  the solutions to outgoing solutions of the free wave equations these solutions determine the electric field everywhere.

\begin{figure}[hbt]
\begin{center}
\includegraphics[width=0.6\linewidth]{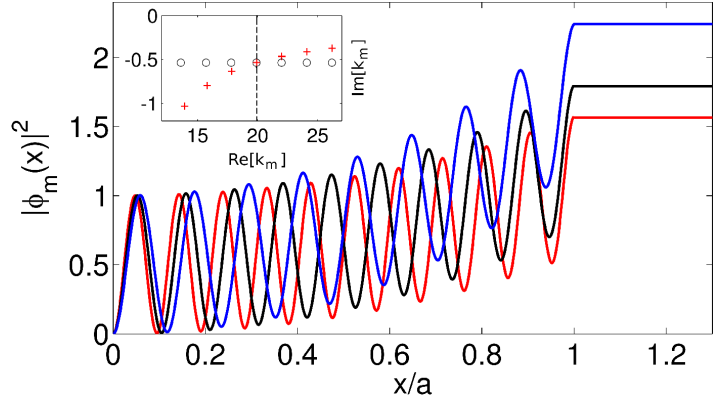}
\caption {\small  Constant-flux (CF) states for simple edge-emitting 1D laser (inset, figure 4, left panel), which are the natural basis set in which to express the lasing solutions of the AISC equation (\ref{eqTSG}).  CF states satisfy the linear wave equation (\ref{eqcf}) with the non-hermitian boundary conditions of only outgoing plane waves at the lasing frequency outside the cavity boundary. Hence for this case they take the form $ \sin (nk_mx)$ with complex amplifying wavevector $k_m$ (3 examples shown).  The complex CF wavevectors are shown in the inset  (red plus signs) and compared to the quasibound wavevectors (black open circles) for $k_\mu = 20$ (dashed vertical line).  Note that the QB and CF wavevectors nearest to the lasing frequency coincide, but otherwise they disagree significantly.} 
\label{fig:cfstates1d}
\end{center}
\end{figure}

Below we will illustrate the theory first with an application to a 1D cavity consisting of a perfect mirror at the origin,
and  a uniform dielectric region of real index $n=n_0$ and length $a$, terminated with vacuum out to infinity in the positive x-direction. For this special case (see the inset of figure~\ref{fig:1dcomp1} left panel) the outgoing wave boundary condition is just $\partial_x \varphi_m (x)|_a = +ik \varphi_m (a)$.  Note this differs subtly but importantly from the quasi-bound state boundary condition, for which the complex eigenvalue $k_m$ replaces the real wavevector $k$ \cite{Tureci06}.  Thus the quasi-bound (QB) states, which are often thought of as becoming lasing modes  when gain is added, are not an appropriate basis, since they have complex wavevectors outside the cavity, do not conserve energy flux there, and diverge at infinity.  This is not true for the CF states, as already noted, which conserve flux outside the cavity, but play the role of the linear cavity resonances inside  the cavity.  Note that the CF and QB states never have {\it exactly} the same complex eigenvalues, and, unlike the QB states, the CF states are a parametric {\it family} of basis functions, parameterized by the lasing frequency, $k$.  When the cavity is very high-Q, there will be a lasing mode near the longest lived cavity resonance (under the gain curve), and in this case there will be one CF state with nearly the same complex eigenvalue (see figure~\ref{fig:cfstates1d}, inset).

Because the CF states are determined by a non-hermitian boundary condition, they are not orthogonal to one another, but instead they are biorthogonal to a set of dual functions, $\bar{\varphi}_{m}(x,k)$, which satisfy the complex conjugate differential equation with the complex conjugate boundary conditions (for the 1D case: $\partial_x
\bar{\varphi}_m (x)|_a = -ik \bar{\varphi}_m (a)$).
In general these dual sets of functions satisfy the biorthogonality relations  \cite{morse&feshbach}:  $\int_{{\cal D}} \, d {\bx}
\, \bar{\varphi}^*_m(\bx,k)  \varphi_n(\bx,k) =  \delta_{mn}$, and are also complete.  These relations make it possible to expand an arbitrary lasing solution,
\begin{equation}
\Psi_\mu (\bx) = \sum_{m=1}^\infty a_m^\mu \varphi_m^{\mu} (\bx)\,,
\label{eqmpansatz}
\end{equation}
so that each $\Psi_\mu$ is defined by this vector of complex coefficients in the
space of biorthogonal functions. Because only CF states with frequencies near the center of the gain curve contribute to the lasing state, it is possible to truncate the sum in (\ref{eqmpansatz}) to a finite number ($N$) of components, making $\Psi_\mu$ a finite dimensional vector.  In the calculations we show here $ 10 \leq N  \leq 25$ is sufficient for convergence (in general higher $D_0$ values require larger $N$).
Recall that the CF boundary condition depends on the lasing frequency, $k_\mu$, so that $\varphi^\mu_m (\bx) =
\varphi_m(\bx, k_\mu)$ and in what follows we define $k^{\mu}_m=k_m(k_\mu)$. By substitution of \eref{eqmpansatz} into (\ref{eqTSG}) and use of the biorthogonality relations one finds:
\be
a^\mu_m = \frac{i D_0 \g}{(\g - i (k_\mu-k_a))}\frac{(k_\mu^2/k_a^2)}{ (k_\mu^2 - k_m^{\mu\,2})} \int_{{\cal D}}d {\bxp}
\frac{ (1+d_0(\bxp))\bar{\varphi}_m^{\mu*} (\bxp) \sum_p a^\mu_p
\varphi^\mu_p(\bxp)
}{\eps(\bxp)(1 +  \sum_{\nu} \Gamma (k_\nu) |\Psi_\nu(\bxp)|^2)}\,.
\label{eqam}
\ee
This is the form of (\ref{eqTSG}) that is employed in our algorithm for finding the lasing modes and frequencies.  As discussed above, it reduces the problem to finding the complex vector of coefficients $\bm{a}^{\mu}$ and the frequency $k_\mu$ for each lasing mode, which depends non-linearly on all the other lasing modes and itself through the infinite order non-linearity evident in the denominator of equations (\ref{eqTSG}), (\ref{eqam}).  

\section{Solution method for AISC equations}

The non-linear amplitude equations in (\ref{eqam}) can be written in the compact form
\begin{equation}
\bm{a}^{\mu} = D_0 \, \bm{T}^\mu \bm{a}^{\mu}
\label{eqTmatrix}
\end{equation}
where $(\bm{T}^\mu)_{mn} \equiv T_{mn} (k_\mu; \{ k_\nu, \bm{a}^{\nu}\})$ is a non-linear operator acting on the properly truncated $N$-dimensional vector space of complex amplitudes $\bm{a}^\mu = (a_1^\mu, a_2^\mu, \ldots, a_{N}^\mu)$ which takes the form:  
\begin{equation}
T_{mn} (k; \{k_\nu, \bm{a}^{\nu}\}) = \Lambda_m (k) \int_{\cal D}  d {\bxp} \frac{ (1 + d_0(\bxp))\bar{\vp}^*_m (\bxp,k)
\vp_n (\bxp,k)}{\epsilon(\bxp) (1 + \sum_\nu \Gamma_\nu |\Psi_\nu (\bxp)|^2)},
\end{equation}
where $\Lambda_m (k) = i \g (k^2/k_a^2)/ [ (\g - i (k - k_a)) (k^2 - k_m^2(k)) ]$. Below some finite value of the pump $D_0$ only the trivial solution exists: $\Psi_\mu = 0 , \forall\mu$. As $D_0$ is increased, a series of thresholds are reached at which the number of non-trivial solutions increases by one.  If we denote the threshold for $i$-mode lasing by $D_{th}^{(i)}$ there exist $i$ solutions $\{k_\mu, \bm{a}^\mu \}$ ($\mu=1,\ldots,i$) for pump parameter $D_0$ such that  $D_{th}^{(i)}<D_0<D_{th}^{(i+1)}$.  In each of these intervals we assume that $i$ solutions to (\ref{eqTSG}) exist and find them by a method to be described below.

\begin{figure}[hbt]
\begin{center}
\includegraphics[width=0.6\linewidth]{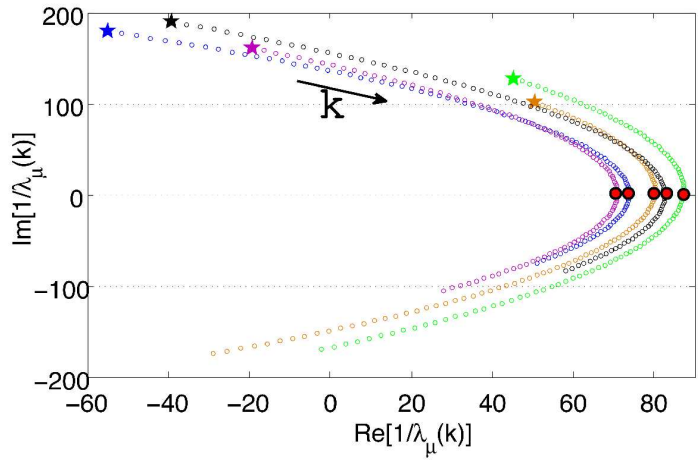}
\caption {\small Eigenvalue flow for the non-interacting threshold matrix $\bm{{\cal T}}^{(0)}(k)$. Coloured circles represent the trajectories in the complex plane of $1/\lambda_\mu(k)$ as a function of $k$ for a scan $k=(k_a-\gamma_\perp,k_a+\gamma_\perp)$ ($k_aR=30,\gamma_\perp R=1$). Shown are only a few eigenvalues out of $N=16$ CF states in the basis. 
The direction of flow as $k$ is increased is indicated by the arrow. Star-shapes mark the initial values of  $1/\lambda_\mu(k)$ at $k=k_a - \gamma_\perp$. The full red circles mark the values at which the
trajectories intersect the real axis, each at a different value of $k=k_\mu$, defining the non-interacting thresholds $D_{th,0}^{(\mu)}$.} 
\label{figT0flow}
\end{center}
\end{figure}

Very close to the first threshold $D_0 = D_{th}^{(1)} + \eps$, we may consider the linearized operator  $\bm{{\cal T}}^{(0)}(k) \equiv \bm{T} (k;\{k_\nu, \bm{a}^\nu = 0\})$
\begin{equation}
{\cal T}^{(0)}_{mn} (k) =  \Lambda_m (k) \int_{\cal D}  d {\bxp} \frac{ (1 + d_0(\bxp))\bar{\vp}^*_m (\bxp,k)
\vp_n (\bxp,k)}{\epsilon(\bxp)} 
\label{eqT0matrix}
\end{equation}
which is obtained by neglecting the $\sum_\nu \Gamma_\nu |\Psi_\nu (\bxp)|^2$ term in the denominator of (\ref{eqTmatrix}).  The resulting linear equation
associated with (\ref{eqT0matrix}) has the form,
\begin{equation}
\bm{{\cal T}}^{(0)}(k) \bm{a}^\mu = (1/D_0) \bm{a}^\mu\,.
\label{eqT0eval1}
\end{equation} 
This equation can in general not be satisfied for a real $D_0$ except at discrete values of $k$.  
$\bm{{\cal T}}^{(0)} (k)$ is a non-hermitian matrix and has $N$ complex eigenvalues $\lambda_n(k)$ for general values of $k$. As $k$ is is varied, the eigenvalues $\lambda_n(k)$ flow in the complex plane, each one crossing the positive real axis at a specific $k_n$, determined by $\mbox{Im}[\lambda_n (k=k_n)] = 0$ (see figure~\ref{figT0flow}). The modulus of the eigenvalue defines the ''non-interacting'' lasing threshold corresponding to that eigenvalue, $D_{th,0}^{(n)} = 1/\lambda_n (k_n)$ (these real eigenvalues will be denoted by $\lambda_\mu^{(0)} = \lambda_\mu^{(0)}(k=k_\mu)$), the real wavevector $k_n$ is the non-interacting lasing frequency, and the eigenvector $\bm{a}^n$ gives the ``direction'' of the lasing solution in the space of CF states. Among these solutions, the smallest $D_{th,0}^{(1)}$ (i.e., the largest of the real eigenvalues $\lambda_\mu^{(0)}$) gives the actual threshold for the first lasing mode; the frequency $k_1$ is the lasing frequency at threshold and the  eigenvector $\bm{a}^1$ defines ''direction'' of the lasing solution at threshold.
The ''length'' of $\bm{a}^1$ cannot be determined from the linear equation (\ref{eqT0eval1}) but rises continuously from zero at threshold and is determined by the non-linear equation~(\ref{eqTmatrix}) infinitesimally above threshold.  As noted, the remaining real eigenvalues of $\bm{{\cal T}}^{(0)}(k)$ define the {\it non-interacting thresholds} for other modes, however the actual thresholds of all higher modes will differ substantially from their non-interacting values due to the non-linear term in (\ref{eqTmatrix}) which now comes into play. The actual lasing frequencies of higher modes have a relatively weak dependence on $D_0$ and differ little from their non-interacting values. Above the threshold $D_{th}^{(1)}$ we solve the non-linear equation (\ref{eqTmatrix}) by an iterative method to be described below.  

\begin{figure}[hbt]
\begin{center}
\includegraphics[width=1.0\linewidth]{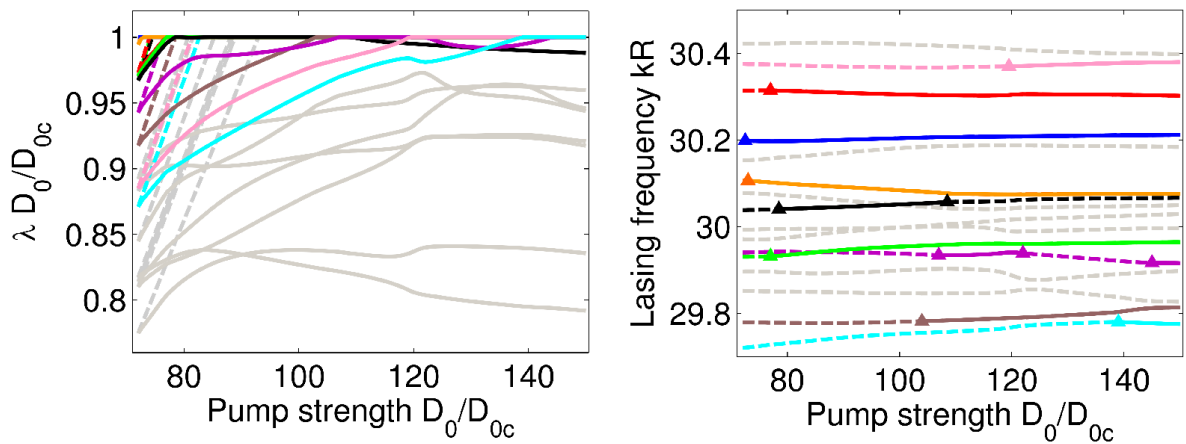}
\caption {\small Evolution of the thresholds and lasing frequencies as a function of $D_0$ for the random laser discussed below (see figure~\ref{fig:drl1}, same colour coding is used). Left: Evolution of the quantities $\lambda_\mu^{(i)} D_0(i)$ as the pump is increased. $\lambda_\mu^{(i)}$ denotes the real $\lambda_\mu$
calculated at step $i$ of the discretization of the full $D_0$-range.  The lasing threshold corresponds to the line $\lambda_\mu D_0=1$, so values below that are below threshold at that pump. Once a mode starts to lase its
corresponding value $\lambda_\mu D_0$ is clamped at $\lambda_\mu
D_0/D_{0c}=1$. The full coloured lines represent the modes which start lasing
within the calculated range of $D_0$. The dashed lines represent the
linear variation $\lambda_\mu^{(0)} D_0(i)/D_{0c}$ of the non-interacting thresholds; gray lines indicate modes which
do not reach threshold in this pump range. Right: Analogous evolution of lasing
frequencies $k_\mu$ with pump. The non-lasing modes are drawn in dashed lines, turning into full lines
as the modes begin to lase. Gray lines represent modes which never
lase in the calculated range of $D_0$.  Note that frequencies of two lasing modes cannot cross (see discussion below), but non-lasing mode frequencies can cross. } 
\label{figmonitorthresh}
\end{center}
\end{figure}

Assuming we have the non-linear solution for the first lasing mode, $\Psi_1(\bx)$, available at each value of $D_0$ (we will denote these discrete values by $D_0(i)$) we can construct the {\it first interacting threshold matrix} as we increment $D_0>D_{th}^{(1)}$:
\begin{equation}
{\cal T}^{(1)}_{mn} (k) =  \Lambda_m (k) \int_{\cal D}  d {\bxp} \frac{ (1 + d_0(\bxp))\bar{\vp}^*_m (\bxp,k)
\vp_n (\bxp,k)}{\epsilon(\bxp) (1 +  \Gamma (k_1) |\Psi_1 (\bxp)|^2)}\,,
\label{eqT1matrix} 
\end{equation}
the second largest eigenvalue of which will determine the second interacting threshold, $D_{th}^{(2)}$. This procedure can be generalized as additional modes turn on to define the second, third, etc.~interacting threshold matrices. This procedure gives us a way to monitor when (\ref{eqTmatrix}) has a second, third, etc.~non-trivial solution. We diagonalize the matrix (\ref{eqT1matrix}) and plot $D_0 (i) \lambda_\mu^{(i)}$ as we increase $D_0$ in small increments solving the non-linear equation (\ref{eqTmatrix}). The resulting plot is shown in figure~\ref{figmonitorthresh}. When $D_0 (i) \lambda_\mu^{(i)}$ reaches unity, a new mode has reached threshold. One can confirm that the $n^{th}$ threshold matrix has $n$ eigenvalues equal to unity until threshold, when the $(n+1)^{th}$ appears. Such a diagram is very interesting because it shows the strong effects of mode competition.  
The eigenvalues of the non-interacting threshold matrix are fixed numbers, independent of pump, $D_0$, so that a 
plot $D_0 \lambda_\mu^{(0)}$ vs.~$D_0$ just gives straight lines intersecting unity at the non-interacting thresholds. The interacting threshold matrix depends strongly on $D_0$, so that the interacting eigenvalues will be sub-linear, leading to much higher thresholds, and some will even be decreasing with increasing $D_0$, indicating modes which are completely suppressed by mode competition and will never turn on.  Thus the interacting threshold matrices give us access to the mode competition {\it below} threshold, as well as the lasing frequencies below threshold, even though the modal amplitudes are strictly zero below threshold in the semiclassical theory.

Using the information obtained by these successive linearizations of the lasing map (\ref{eqTmatrix}) we can begin our iteration process of the non-linear problem with a very accurate approximation to the lasing frequencies and the vectors $\bm{a}^\mu$ defining each lasing mode.  This allows the iterative solution to converge in a reasonable time.  However there is one additional subtlety concerning (\ref{eqTmatrix}); this equation is invariant under a global phase change $\bm{a}^\mu \rightarrow \ex{i\phi} \bm{a}^\mu$ .  Each member of this continuous family of solutions should correspond to a real non-linear eigenvalue $1/D_0$ for each value of the pump.  However if we iterate our trial solution the phase of the iterate continuously evolves and never converges, although the amplitudes of the components do converge.  To find a unique solution the global phase is fixed (we refer to this as the ``gauge condition'') and the lasing frequency is allowed to flow on each iteration to maintain this global phase.
This is the analog for the non-linear problem of allowing the eigenvalues of the threshold matrices to flow to the real axis by tuning frequency eigenvalues discussed above, and depicted in figure~\ref{figmonitorthresh}.  It ensures that the non-linear eigenvalue is unity, and not a complex number of modulus unity. In this manner the interacting lasing frequencies can be found above threshold and will differ from the non-interacting or threshold values. In practice, we choose the gauge in which we set $\im{a_{M_\mu}^\mu}=0$, where $M_\mu$ is the largest CF component of the eigenvector $\bm{a}^\mu$ of the non-interacting threshold matrix $\bm{{\cal T}}^{(0)}(k)$.

To summarize our solution procedure for (\ref{eqTmatrix}): the givens are the dielectric function of the resonator, $\eps(\bx)$, the atomic frequency, $k_a$ and gain width $\g$.

\begin{itemize}

\item
The CF states $\varphi_m$ and wavevectors $k_m$ are found for a range of $k$ values near $k_a$ by solving the linear non-hermitian differential equation (\ref{eqcf}).

\item
The first threshold matrix $\bm{{\cal T}}^{(0)}(k)$ is constructed from the CF states (and their adjoint partners) and frequencies following (\ref{eqT0matrix}). $k$ is varied and its largest real eigenvalue is found at $k=k_{max}$;  this defines the first threshold and the lasing frequency $k_1 = k_{max}$; the corresponding eigenvector determines the trial solution just above threshold.

\item
The pump $D_0$ is increased in small steps and the solution from the previous step is used as the trial solution for the next step to solve the non-linear equation (\ref{eqTmatrix}) efficiently by iteration. The lasing frequencies are allowed to flow to maintain the gauge condition as discussed above.

\item
The solution for the non-linear equation with $n$ lasing modes is used to construct the $n^{th}$ interacting threshold matrix,  $\bm{{\cal T}}^{(n)}(k)$, which is continuously monitored for the appearance of an additional eigenvalue $D_0 \lambda(k)=1$, signaling the threshold for mode $(n+1)$, and the need to increase the number of solutions to the non-linear equation (\ref{eqTmatrix}).

\end{itemize}

Codes based on this algorithmic structure have been developed successfully for arbitrary 1D cavities, for 2D uniform dielectric cavities of general shape, and for 2D disordered cavities embedded in a disk-shaped gain medium.

\section{Solution of 1D cavity and comparison to time-dependent solutions}

The advantage of our new approach to semiclassical laser theory is that it gives a time-independent theory of the stationary lasing states, which directly calculates the physical quantities of interest (thresholds, lasing frequencies, electric fields inside the cavity and outside). The method is computationally much more efficient than fully time-dependent simulations, and it provides a physical picture of the lasing solutions as well as analytic results near thresholds.  Previous time-independent approaches were not quantitative and didn't treat the openness of the system fully.  Our method is based on one key approximation, that of stationary inversion, which
holds when $\gp/\g, \gp/\Delta k_{\mu \nu} \ll 1$, where $\gp$ is the relaxation rate of the inversion and $\Delta k_{\mu \nu}$ is the frequency spacing of any two lasing modes.  We tested the accuracy of our approach recently  \cite{LiTST08} by comparing our AISC results for a simple 1D edge-emitting laser with exact time-dependent simulations of the Maxwell-Bloch (MB) equations for the same system (no assumption of stationary inversion).  In the latter case the system was simulated until steady-state was reached, and then the resulting oscillatory fields were Fourier-analyzed to yield modal intensities and frequencies to compare with the quantities $\psimu$, $k_\mu$  found with the AISC laser theory.

\begin{figure}[hbt]
\begin{center}
\includegraphics[width=0.9\linewidth]{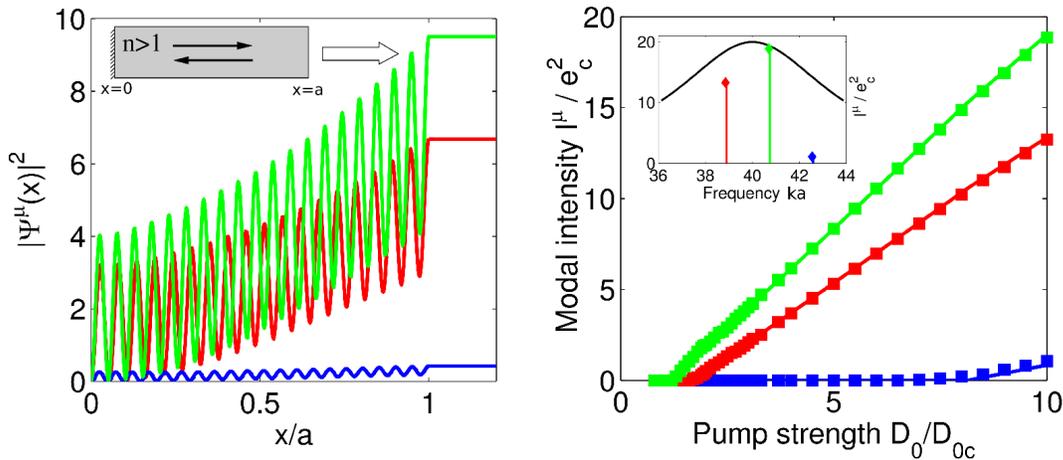}
\caption {{\small Left: Results of AISC calculation for 1D edge-emitting laser with perfect mirror at the origin and a dielectric step ($ n = 1.5 \to n=1$) at $x=a$ (inset). At this pump level, $D_0 = 10$, three modes are lasing with different frequencies and different spatial structures. It is apparent from their growth to the loss boundary that these modes do not conserve photon flux and are not the solution of a hermitian problem inside the cavity, although they do conserve flux outside $(x/a > 1)$.  Right: Data points are MB time-dependent solutions Fourier-analyzed at stationary output power, solid lines are results of AISC calculation requiring $\sim 10^{-3}$ times the computational effort. 
Color coding is the same as for the left panel. Inset shows lasing frequencies compared using the two approaches (diamond vs.~solid coloured lines); black curve is the gain profile.}} 
\label{fig:1dcomp1}
\end{center}
\end{figure}

For the 1D edge-emitting laser of length $a$, uniform index, $n_0$, with perfect mirror at the origin (see figure~\ref{fig:1dcomp1}, left panel) the solutions for the CF states are simply,  
\begin{equation}
\varphi_m (x,k) = \left\{\begin{array}{lcc}
\eta_m (k) \sin (n_0 k_m (k) x)&{\rm for}&x<a \\
\eta_m(k) \sin(n_0 k_m (k) a)\ex{i k (x-a)}&{\rm for}&x>a \end{array}\right.\,,
\end{equation}
where $\eta_m (k)$ is a $k$-dependent normalization constant chosen to give $\delta_{mn}$ in the biorthogonality relation.
The complex wavevectors $k_m (k)$ are found through the characteristic equation,
\begin{equation}
\tan(nk_m a) = -i\frac{nk_m}{k}\,,
\label{eqLABsec1d}
\end{equation}
and always have a non-zero negative imaginary part, corresponding to amplification from left to right. The lasing solutions are found by the procedure outlined in the section above.  A single CF state is not an adequate solution for $n_0 = 1.5$, but typically only contributions from three CF states are needed and the first lasing mode has primary contributions from the CF states closest to the atomic frequency, $k_a$.  In figure~\ref{fig:1dcomp1}  we show typical results of the AISC laser theory, as well as a comparison of AISC and MB solutions. Excellent agreement is found with no free parameters, well into the regime of multi-mode lasing in which the inversion is not strictly stationary. This is the first demonstration, to our knowledge, of successful quantitative agreement between solutions obtained by time-independent methods and the full time-dependent solution of MB equations.

In \cite{LiTST08} we varied the parameters $\gp$, $n_0$ and good agreement was found over a wide range; in addition, deviations when $\gp \to \g$ were shown to be amenable to perturbative treatment. The time-dependent solutions were full-wave Maxwell-Bloch, because we found that the slowly-varying envelope approximation was  not very accurate \cite{LiTST08}; hence these simulations  were required to cover the large difference in time scales between rapid oscillations at $k_a^{-1}$ and the slow equilibration at $\gp^{-1}$, making them very time-consuming. In contrast, our theory is more accurate as $\gp^{-1}$ increases compared to other dynamical time scales and requires no more computational effort.  Therefore we believe our method may be the only practical numerical technique for solving the MB equations for more complex and more realistic three-dimensional laser structures.  In addition, our method is infinite order in the non-linearity and we found \cite{LiTST08} that a 3rd order treatment of the non-linear interactions failed badly.  

\begin{figure}[hbt]
\begin{center}
\includegraphics[width=0.9\linewidth]{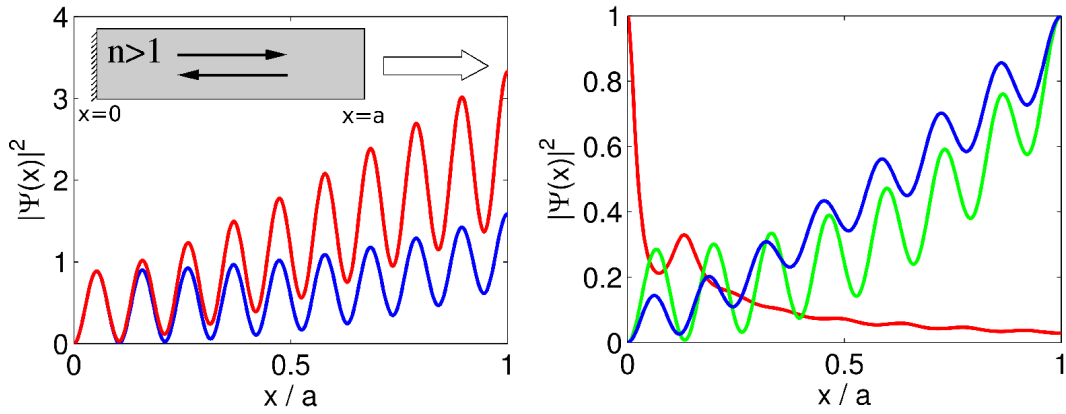}
\caption {{\small Left: The spatial profile of a lasing mode just above threshold (blue curve) and well above threshold (red curve). Non-linear self-interaction has mixed in higher CF states, increasing the amplification rate for this mode.   The gain width $\gamma_\perp =0.5 $ was chosen to maintain single-mode lasing over wide pump range.
Right: Effect of spatial hole burning interaction on higher modes: Red curve is the stationary inversion well above threshold in the multi-mode regime for the 1D laser (inset).  Blue curve is the second lasing mode above threshold, compared to its profile (green) in the absence of interactions.  Note the reduction in spatial oscillations due to interactions. The results are scaled so that the maximum value is unity. }} 
\label{fig:1dcomp2}
\end{center}
\end{figure}

As the lasing solutions evolve above threshold, they interact with themselves and with other modes in a space-dependent manner.  The electric fields burn a hole in the inversion which acts as a (complex) index variation for the other modes, and for itself.  As a consequence the ''shape'' of the modes continues to change substantially with pump. This effect is neglected in conventional approaches which assume each lasing mode is a single fixed cavity mode, with a varying overall scale factor. figure~\ref{fig:1dcomp2} illustrates this effect.

\section{Diffusive random lasers}

The diffusive random laser (DRL) is perhaps the most challenging new system of interest for fundamental laser physics. In its most basic realization, it consists of a random aggregate of particles which scatter light and have gain or are embedded in a background medium with gain [12, 15--23, 25--27]. 
The system has no traditional resonator to trap the light;  photons generated by stimulated emission are only very minimally ''confined'' by multiple scattering as it diffuses out of the medium.  The diffusive escape is so rapid that {\it the medium exhibits no isolated resonances at all} in the absence of gain.  Standard approaches to multi-mode laser theory assume each lasing mode is described spatially by a linear cavity mode, usually a closed cavity mode, and thus were unsuited to deal with such an extremely leaky cavity.  Despite the lack of sharp resonances, the laser emission from DRLs was observed to have the essential  properties of conventional lasers: the appearance of coherent emission with line-narrowing above a series of thresholds, and uncorrelated photon statistics above threshold indicative of gain saturation \cite{Cao05,CaoZHSWC99,CaoLXCK01,Cao03}. Earlier numerical work (for a recent review see \cite{Wiersma08}) has suggested: 1) That the lasing modes are coherent, and not the result of incoherent amplification as originally supposed. 2) That these modes exist outside of the strong disorder regime, where high-Q linear modes arise due to Anderson localization. These observations raise the obvious questions: what are the nature of the lasing modes in DRLs?  What determines the lasing frequencies, which are not determined by long-lived cavity resonances?  What is the effect of modal interactions in such complex and low finesse cavities?

\begin{figure}[hbt]
\begin{center}
\includegraphics[width=0.7\linewidth]{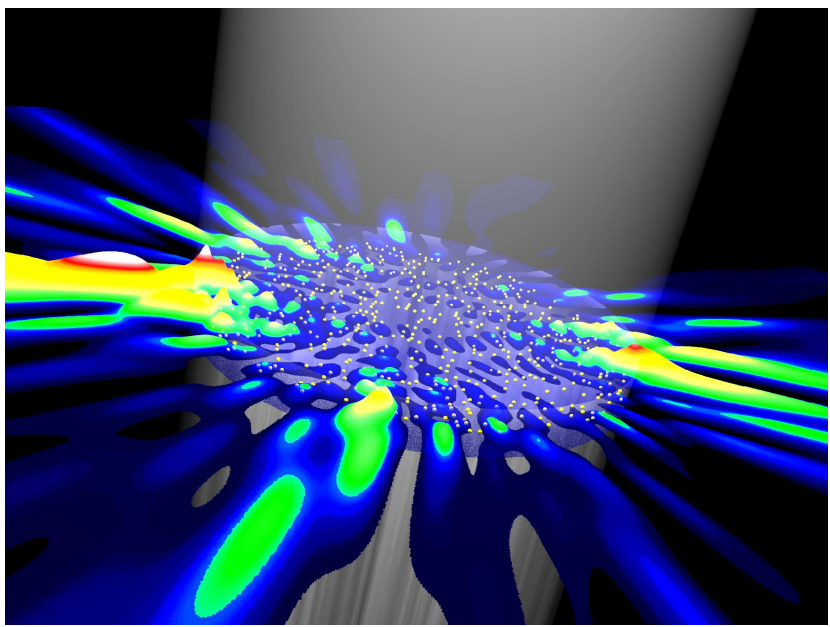}
\caption{\small Three dimensional rendering of the electric field from AISC
numerical calculations in the random laser discussed below (see figure~\ref{fig:drl1}), in which the yellow spheres represent the nano-particles in a cylindrically symmetric gain medium. The electric field, where color and height indicate intensity, represents the steady state solution of the Maxwell-Bloch equations at $D_0/D_{0c}=123.5$. The system is illuminated uniformly by incoherent light, shown in this figure as coming from above.}
\label{fig:cover}
\end{center}
\end{figure}

\begin{figure}[hbt]
\begin{center}
\includegraphics[width=0.6\linewidth]{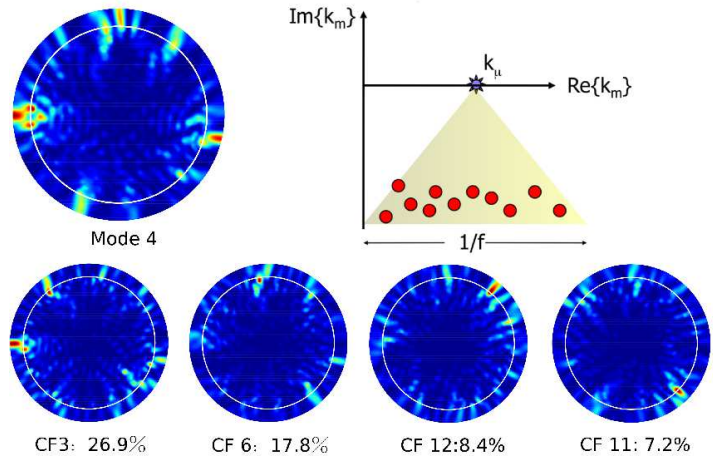}
\caption{\small Top left: false colour intensity plot of typical mode of DRL; example is mode 3 at $D_0/D_{0c} = 123$ in figure~\ref{fig:drl1}.  Bottom: four CF states which make the largest contribution to the lasing mode. White line is boundary of disk-shaped gain region. Note that while the CF states fluctuate in space, their variance is largest at the cavity boundary.  This is due to their non-hermitian nature and is discussed below. Top right: schematic illustrating composite nature of DRL lasing modes.  One mode with a real frequency $k_\mu$ has contributions from roughly $1/f$ CF states with complex wavevector of the type shown in the bottom panels.}  
\label{fig:cfstates}
\end{center}
\end{figure}

The AISC theory is the first time-independent approach which can solve the semiclassical lasing equations for a multi-mode DRL; a typical example for a calculation of the multi-mode electric field both inside the ``cavity" and outside is rendered in three-dimensions in figure~\ref{fig:cover}.  Our first results 
 \cite{TureciGRS08} provide a framework for thinking about the lasing modes in DRLs.  In \cite{TureciGRS08} we modeled a 2D DRL by a disk-shaped gain medium which contains a random distribution of weakly scattering sub-wavelength particles. Such a system has strongly-overlapping linear resonances; in the language of resonator theory, its finesse, $f$, (the ratio of the resonance spacing on the real axis to the typical distance of a resonance from the real axis) is much less than unity. The AISC laser theory indicates that in such a situation the lasing modes will be a superposition of many CF states with roughly equal weight; an example of this is shown in figure~\ref{fig:cfstates}. Thus, as compared to the edge-emitting laser, a larger number of modes $N \sim 1/f$ have to be kept in the expansion (\ref{eqmpansatz}). The CF states of a DRL around a given $k_a$ have comparable decay rates $\kappa_m \equiv \im{k_m}$ and a similar statistical distribution as the resonances \cite{TureciGRS08}. 

\begin{figure}[hbt]
\begin{center}
\includegraphics[width=0.6\linewidth]{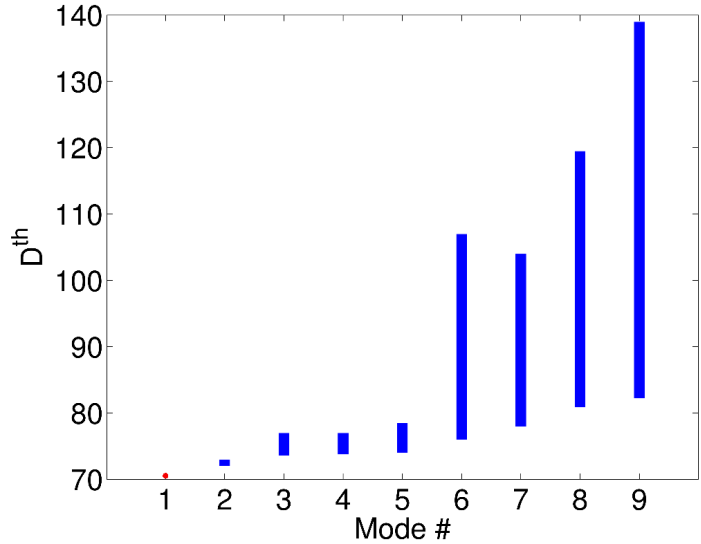}
\caption{\small The shift of the exact thresholds from their non-interacting values. The bottom end of the blue lines denote the thresholds determined from the eigenvalue flow of $\bm{{\cal T}}^{(0)}(k)$ while the top of the lines mark the exact thresholds calculated by solving (\ref{eqTSG}). The red dot at mode number 1 indicates that this threshold does not shift.}  
\label{fig:threshift}
\end{center}
\end{figure}

Large DRLs with $k_aR \gg 1$ ($R$ is the linear scale of the random aggregate) are typically highly multi-mode laser oscillators at moderate pumping strengths, due to the existence of many CF states with similar ``decay rates", $\kappa_m$.  As a consequence, modal interactions through spatial hole burning are extremely strong and have to be taken into account correctly. This manifests itself first and foremost in the large shifts of the interacting thresholds from their non-interacting values calculated from $\bm{{\cal T}}^{(0)}(k)$, as shown in figures~\ref{fig:threshift} and \ref{figmonitorthresh} (above).

\begin{figure}[hbt]
\begin{center}
\includegraphics[width=0.8\linewidth]{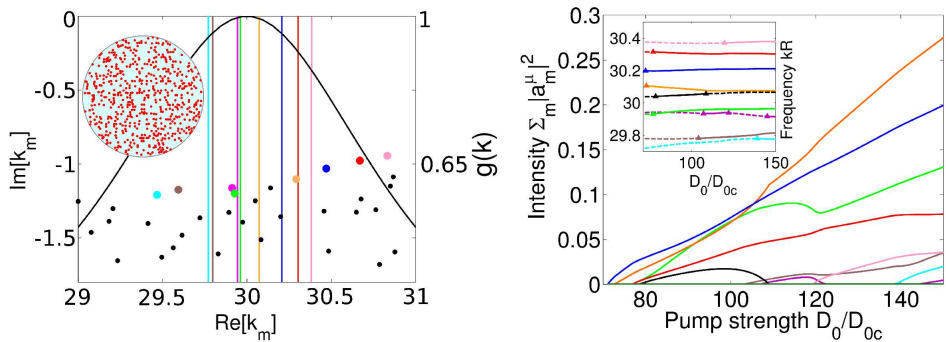}
\caption {{\small Left: Frequencies of multi-mode DRL well above threshold (coloured lines).  Filled circles are complex frequencies of the CF states contributing to lasing modes; coloured circles denote state with largest contribution to each mode, but in general each mode is a superposition of many CF states. Solid black curve is the gain profile. Inset:  DRL model with a random configuration of scatterers in uniform disk of gain medium. Right: Modal intensities vs.~pump for same DRL. Note complex non-monotonic behaviour due to modal interactions.  Inset: lasing frequencies  vs.~pump above (solid) and below (dashed) threshold.}} 
\label{fig:drl1}
\end{center}
\end{figure}

The frequency $k_\mu$ of a lasing line for a DRL is collectively determined. In \cite{TureciGRS08} we derived a relation of the form:
\be
k_\mu \approx
k^{(0)}_\mu  + k^{(c)}_\mu
\label{eqkmu}
\ee
where $k^{(0)}_\mu $ satisfies the standard relation for the frequency of a single-mode laser \cite{haken_light2}, which is a weighted average of the cavity and atomic transition frequencies (we take the largest CF component at threshold as the cavity frequency). $k^{(c)}_\mu$ is a collective contribution due to all the other CF components which has no analog in conventional lasers and involves a sum over the off-diagonal elements of $T^\mu_{mn}$.
When $k_\mu R \gg 1$ the collective term dominates and we expect lasing frequencies to be distributed randomly (although correlated) under the gain curve.  This is roughly what was found (see figure~\ref{fig:drl1}).

The evolution of intensities in a DRL as a function of pumping strength $D_0$ (figure~\ref{fig:drl1}, right panel)  is very different from that of a conventional edge-emitting laser (figure~\ref{fig:1dcomp1}). The intensity variation displays a complex non-monotonic behaviour due to strong modal interactions. This is simultaneously reflected in the evolution of the lasing frequencies, illustrated in the inset (right panel) of figure~\ref{fig:drl1} above.  When two lasing frequencies approach one another, instead of repelling, one of the frequencies disappears (e.g., the ``black'' frequency in figure~\ref{fig:drl1}), because the corresponding mode is driven to zero, an effect which can't happen in a linear system. Recall that these frequencies are not the eigenvalues of a random matrix but rather the parameter values $k_\mu$ which make the random operator $T^\mu_{mn}$ have a real eigenvalue.  It is possible to analyze this condition in terms of the interacting threshold matrices defined above.  In order for frequency degeneracy to occur this random complex matrix would have to have two degenerate real eigenvalues at the same value of $k_\mu$.  We argue \cite{TureciGRS08} that level repulsion in the complex plane prevents this from happening.  However, instead of leading to frequency repulsion, this effect leads to frequency locking, i.e., the two interacting modes merge into  one.  We can confirm this effect by noting that after one modes ``dies", the CF decomposition of the surviving mode has large components associated with the mode which was driven to zero.  So instead of the ``exchange of identity'' familiar from linear level repulsion, we find ``merger of identity''. In a time-dependent treatment, this kind of cooperative frequency locking is well-known, and we studied it in a chaotic cavity laser in \cite{tureci05b}.  Interestingly, our results show that it is even possible for a mode to ''die'' and then to ``reincarnate'' itself when its frequency moves far enough away from that of the dominant mode (see the mode represented by purple line in figure~\ref{fig:drl1}).

Recent experiments \cite{vanderMolenTML07} have shown DRL frequencies to be equally spaced and have compared them to the eigenvalue distribution of the gaussian orthogonal ensemble of random matrices, which our calculations suggest may not be appropriate, since the frequencies are not themselves the eigenvalues of a random matrix.  The same experiments have shown that for a fixed impurity configuration the frequencies are rather stable between different pump pulses, while the modal intensities vary widely.  We expect this behaviour due to the modal interactions.  The frequencies only vary slowly as a function of pump, whereas the intensities vary strongly as seen in figure~\ref{fig:drl1} above. Thus the exact pump strength or spatial profile of the pump would not be expected to change the frequencies much.  However, the interactions can amplify the effect of any small change in lasing threshold of one mode vs.~another. In \cite{TureciGRS08} we found that a 30\% random variation in the pump profile (for fixed impurity configuration) hardly changed the lasing frequencies at all well above threshold, but changed the relative modal intensities by factors of order unity (see figure~\ref{fig10} below).

\begin{figure}[hbt]
\begin{center}
\includegraphics[width=0.9\linewidth]{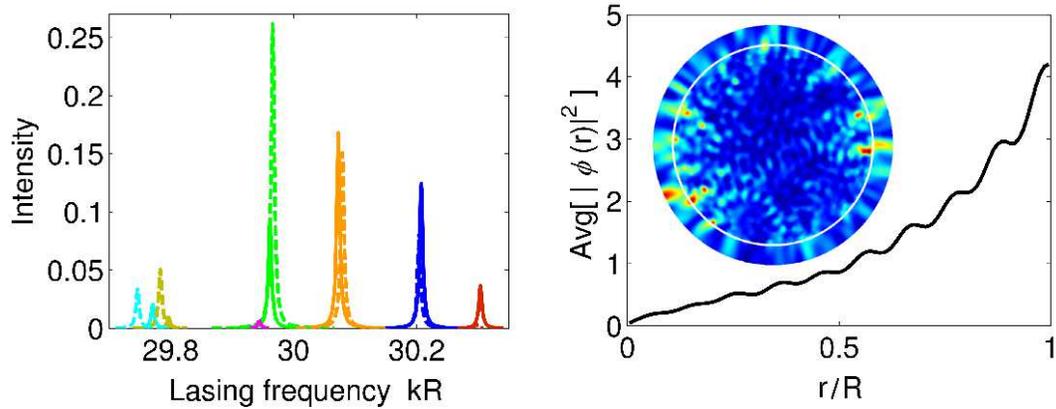}
\caption{{\small Left: Intensity and frequency fluctuations in a DRL. For fixed impurity configuration we compare lasing frequencies and intensities in the multimode regime for uniform pump (solid lines) and uniform pump with 30\% zero mean spatial noise added (dashed lines). Right: Field distribution of a DRL. Inset: False colour plot of electric field intensity of the DRL of figure~\ref{fig:drl1} (seven modes lasing) at pump $D_0/D_{0c} = 123.5$ (white circle is boundary of gain medium). Note brightest regions appear at the edge of the gain medium.  Main plot:  Radial intensity of CF states contributing to the lasing modes averaged over 400 disorder configurations.  There is a large non-random increase of intensity with $r$.}}  
\label{fig10}
\end{center}
\end{figure}

Finally, we consider the spatial variation of the electric field in DRLs (figures~\ref{fig:cfstates}, \ref{fig10}).  The false colour
representation of the multi-mode electric field in the laser has a striking property: it is consistently brighter at the edge of the
disk than at its center, even though the gain is uniform and there are no special high-Q modes
localized near the edge.  To demonstrate that this effect is not a statistical fluctuation associated with this particular disorder
configuration we have averaged the behavior of the entire basis set of CF states over disorder configurations.  The result is an average growth of intensity towards the boundary.  The origin of this is the complex amplifying character of the CF states, which lend the same character to the lasing modes which they form by superposition (compare to figures~\ref{fig:cfstates1d} and \ref{fig:1dcomp1}). The non-hermitian boundary condition causes ``growth to loss'', i.e., a tendency for the electric field to be higher near the loss boundary where it must compensate for photons leaking out in steady-state. The leakier the laser cavity, the stronger this effect is, so it is particularly pronounced in DRLs which have finesse much less than unity.

Note that this effect means
that the electric field fluctuations in DRLs will differ
substantially from the random matrix/quantum chaos fluctuations of
linear cavity modes\cite{stockmann}, first because each mode is a superposition of
pseudo-random CF states and second because these CF states
themselves are not uniform on average.  We are currently studying both the statistical distribution for lasing frequencies and for the spatial characteristics of the lasing modes for the DRL.  Another system of interest in this regard are the dielectric cavity lasers with chaotic ray dynamics \cite{SchwefelTureci04}.  Here the complexity just arises from the shape of the boundary, which need not be highly transmissive, so that the statistical properties of the lasing modes should be close to those of single CF states.  

\section{Summary and conclusions}

A laser is an open non-linear system which self-organizes to exhibit multiperiodic behavior.  In conventional approaches the lasing modes are described by linear cavity resonances and frequencies.  We have recently found a method to determine the steady-state of a laser which does not start from  the linear cavity properties, but allows the laser to self-consistently ``find'' the steady-state lasing modes and frequencies.  These quantities are determined by a set of infinite order non-linear integral equations, for which we have developed an efficient, iterative solution algorithm. This time-independent self-consistent method has been tested against exact numerical solution of the Maxwell-Bloch (MB) lasing equations for a 1D cavity and has been found to provide highly accurate solutions over a wide and physically relevant parameter range.  In our time-independent approach the lasing modes are in general a superposition of non-hermitian linear cavity states, termed constant-flux (CF) states.  The more open the laser cavity, the more terms contribute to this superposition.  An extreme example is the diffusive random laser (DRL), which is so leaky that is has no detectable linear cavity resonances.  We have shown that our method can find the lasing modes of such a system, and reveals the effects of strong modal interactions.  The theory presented here is ``ab initio'' in the sense that it generates all properties of the lasing states from knowledge of the dielectric function of the host medium and two parameters of
the gain medium (the frequency of the lasing transition and its relaxation rate).  We refer to it as ab initio self-consistent (AISC) laser theory.  The method should be applicable to any novel laser-cavity system, and if generalized to treat the gain medium more realistically, may allow accurate simulation of lasers for design and optimization purposes.

The linear CF states provide a link between the problem of quantum/wave chaos and the steady-state lasing properties of disordered or wave-chaotic resonators.  We hope to combine these linear methods along with our non-linear self-consistent equation to provide a statistical description of random or wave-chaotic lasers.  It will also be important to incorporate this full non-linear treatment into a quantum theory of open laser cavities 
\cite{viviescas}, which at present remains a challenging and only partially solved problem.

 \ack
This work was supported by NSF grant DMR-0408636, by the Max Kade and W M Keck foundations and by the
Aspen Center for Physics. We thank Manabu Machida, Braxton Collier, Hui Cao, Ad Lagendijk, Patrick Sebbah, 
Christian Vanneste and Diederik Wiersma for discussions.


\end{document}